\documentclass[twocolumn,aps,pra,floatfix]{revtex4}
\usepackage{graphicx,amssymb,amsmath}
\begin{document}
\title{Anomalous Dimers in Quantum Mixtures near Broad Resonances:\\
Pauli Blocking, Fermi Surface Dynamics and Implications}
\author{Jun-Liang Song}
\affiliation{Institute for Quantum Optics and Quantum Information of the Austria
Academy of Sciences, A-6020 Innsbruck, Austria}
\author{Fei Zhou}
\affiliation{Department of Physics and Astronomy, University of British Columbia, Vancouver, B.C., Canada V6T1Z1}
\date{June 13, 2011}

\begin{abstract}
We study the energetics and dispersion of anomalous dimers that are induced by
the Pauli blocking effect in a quantum Fermi gas of majority atoms near
interspecies resonances.
Unlike in vacuum, we find that both the sign and magnitude of the dimer masses
are tunable via Feshbach resonances.
We also investigate the effects of particle-hole fluctuations on the dispersion
of dimers and demonstrate that the particle-hole fluctuations near a Fermi surface
(with Fermi momentum $\hbar k_F$) generally reduce the effective two-body interactions and the binding energy of dimers.
Furthermore, in the limit of light minority atoms the particle-hole fluctuations disfavor the formation of dimers with a total momentum $\hbar k_F$,
because near $\hbar k_F$ the modes where the dominating
particle-hole fluctuations appear are the softest.
Our calculation suggests that near broad interspecies resonances when the
minority-majority mass ratio $m_B/m_F$ is smaller than a critical value
(estimated to be $0.136$), dimers in a finite-momentum channel are
energetically favored over dimers in the zero-momentum channel.
We apply our theory to quantum gases of $^{6}$Li$^{40}$K, $^{6}$Li$^{87}$Rb,
$^{40}$K$^{87}$Rb and $^{6}$Li$^{23}$Na near broad interspecies resonances, and
discuss the limitations of our calculations and implications.
\end{abstract}

\maketitle

\section{Introduction}
Pairing of fermions or electrons in quantum condensed matter systems for long has been
one of the most remarkable phenomena discovered in many-body
physics \cite{Schrieffer99,Larkin65,Fulde64}. It is also the fundamental issue in
the so-called color-superconductivity phenomenon in the core of a neutron
star \cite{Alford08}. Recent developments in the field of ultracold atoms,
especially the observation of atomic Feshbach resonances, have further led to
tremendous new opportunities of studying novel pairing correlations in quantum
gases \cite{Chin08,Zwierlein06,Partridge06,Liao10}. Because of tunability of
interactions and availability of a rich variety of alkali isotopes, cold gases
have been turned into a marvelous, promising platform to study diversified
pairing phenomena with unique correlations. Pairing properties near
Feshbach resonances are mainly determined by, apart from scattering lengths,
asymmetries in chemical potentials of the different atoms involved. Such asymmetries
and consequently a very rich class of pairing phenomena are known either due to
the imbalance in populations \cite{Zwierlein06, Partridge06, Liao10, Son06, Pao06,
Sheehy06,Gu06,Gubankova03,Bulgac08} or tunable mass ratios \cite{Petrov07,
Baranov08, Gezerlis09, Huse10}, or both. Moreover, differences in the quantum
statistics of underlying atoms also contribute to the asymmetries in chemical
potentials as in Fermi-Bose mixtures \cite{Simoni03, Stan04,Inouye04,Ferlaino06,
Ospelkaus06a,Ospelkaus06b,Deh08,Ni08}.
Previous theoretical studies on Fermi-
Bose mixtures have been focused on the formation of molecular Fermi surfaces
near narrow resonances \cite{Powell05}
or far away from
resonances \cite{Bortolotti06}, pair correlations \cite{Watanabe08}, and quantum and
thermal depletion of condensates \cite{Frantini10}.

A very closely related fascinating issue that has been challenging to the cold
atom community is whether near-resonance two-body scattering in a channel with
nonzero total momentum dictates the instability in interacting quantum
mixtures. A major challenge in the studies of pairing phenomena in various
quantum mixtures near resonances, unlike in the BCS-BEC crossover regime of a
Fermi gas, is that there are more competing pairing schemes and possibilities of
having higher order correlations; energetically it is quite difficult to
differentiate competing scenarios near resonances. One natural approach is of
course to perform a full-scale numerical simulation to resolve the issue. An
alternative is perhaps to study few-body physics in the presence of a finite-density
quantum gas and to gain insight on many-body correlations by exploring
implications of few-body physics. A few interesting attempts have already been
made along this direction. For instance, the properties of
a single minority atom submerged in a quantum Fermi gas have been thoroughly investigated
as a diagnosis of many-body correlations near interspecies Feshbach resonances
 \cite{Chevy06,Pilati08,Combescot07,Prokof'ev08,Punk09,Mora09,Bruun08,Schirotzek09}.
For Fermi-Bose mixtures, anomalous dimers with
tunable masses were emphasized and the leading effect of particle-hole
fluctuations had been studied diagrammatically \cite{Song10}. At the same time,
for Fermi-Fermi mixtures various instability lines have been determined
numerically by studying the dimer and trimer formation in a truncated Hilbert
space and the role of anomalous dimers as well as a universal trimer was
explored \cite{Huse10}. And very recently, Efimov states in a Fermi gas were
studied in a few limiting cases and the spectrum flow has been obtained in the
static-Fermi-sea approximation \cite{MacNeill10}. Logically speaking, these
studies should form potential building blocks for constructing many-body states;
they can also serve as a starting point for more systematic studies on
the interplay between few- and many-body physics in cold gases near resonances.

 In this paper, we take a further step along this direction hoping that our
 efforts to understand dressed bound states can shed more light on many-body
 pairing phenomena in mixtures. Particularly, we consider the limit of {\it a
 single atom or a minority atom in resonance with majority Fermi atoms which
 form a Fermi sea.} This minority atom can be either a fermion or boson although
 most of our discussions are in the context of a Bose atom resonating with a
 Fermi sea. We obtain the dispersion of dimers and investigate the Pauli
 blocking effect, the effect of fluctuating particle-hole pairs near the Fermi
 surface on the dimer energetics. These results should be applicable to quantum
 mixtures with extremely imbalanced population. We further examine the dimer
 dispersion near broad interspecies resonances for $^{6}$Li-$^{87}$Rb \cite{Li08,Marzok09},
 $^{6}$Li-$^{23}$Na \cite{Stan04,Gacesa08}, $^{40}$K-$^{87}$Rb \cite{Simoni03, Inouye04, Ferlaino06,
 Ospelkaus06a, Ospelkaus06b,Deh08, Ni08} and
 $^{6}$Li-$^{40}$K \cite{Dieckmann08,Innsbruck08,Walraven10} that have been available in laboratories.

Our analysis on the dispersion of dimers further suggests that when the minority
atom is very light, minority-majority atoms near broad resonances would prefer
to form dimers in finite-momentum channels that are energetically favorable.
Although quantum mixtures of $^{6}$Li-$^{40}$K, $^{6}$Li-$^{87}$Rb, $^{40}$K-$^{87}$Rb
and $^{6}$Li-$^{23}$Na so far studied in laboratories are not in this particular
limit, we anticipate such a limit can be explored in future generations of
quantum mixtures. More importantly, when combining the optical lattices and
Feshbach resonances \cite{Ospelkaus06b,Best09}, one can achieve resonances with
continuously tunable widths and locations \cite{Cui10}. By properly
choosing laser intensities, or by using isotope selective optical
lattices \cite{Best09}, the ratio of band masses of Fermi-Bose atoms can be
further continuously tuned over a very wide range.
This leads to potential opportunities of studying the limit of light Bose atom.

The rest of the paper is organized as follows.
In Sec. II, we introduce the model Hamiltonian for
broad resonances and study the formation of dimers in a static Fermi sea.
We analyze the effects of Pauli blocking on the dispersion of dimers,
and illustrate that a kinematic effect in the limit of very light minority atoms
favors finite-momentum dimers near broad resonances.
We then present the results for
quantum mixtures so far studied in laboratories.
In Sec. III, we go beyond the static-Fermi-sea approximation
to take into account various corrections due to particle-hole fluctuations
near the Fermi surface of majority fermions.
Our calculations indicate that the main effect of fluctuating pairs is to
reduce the binding energy of dimers and also to
disfavor the formation of dimers with total momenta close to $\hbar k_F$.
In Sec. IV, we discuss the limitations of our results,
explore the implications
on near-resonance quantum mixtures with a finite density of minority atoms
and comment on
various results \cite{Song10,Petrov07,Bulgac08,Huse10,MacNeill10} obtained in previous studies.
We summarize the results and discussions in Sec. V.

\section{Anomalous Dimers: Pauli blocking effect}

We start with a single-channel Hamiltonian
with a short-range interaction $U_{bf}$ that
was introduced previously for a study of Fermi-Bose mixtures
near broad resonances \cite{Song10},
\begin{eqnarray}
&H&=\sum_{\bf k} \epsilon_{\bf k}^F f^\dagger_{\bf k}f_{\bf k}+
\sum_{\bf k} \epsilon_{\bf k}^B b^\dagger_{\bf k}b_{\bf k}
\nonumber \\
&+& \frac{U_{bf}}{\Omega}
\sum_{{\bf k},{\bf k}',{\bf Q}} f^\dagger_{
\frac{m_R}{m_B}
{\bf Q}
+{\bf k}}
b^\dagger_{\frac{m_R}{m_F}{\bf Q}
-{\bf k}}
f_{\frac{m_R}{m_B}
{\bf Q}
+{\bf k}'}b_{
\frac{m_R}{m_F}
{\bf Q}
-{\bf k}'}
\label{Hamiltonian}
\end{eqnarray}
Here $f^\dagger_{\bf k}$,$b^\dagger_{\bf k}$($f_{\bf k}$,$b_{\bf k}$) are
creation (annihilation) operators for Fermi and Bose atoms respectively, and
$\epsilon^{F(B)}_{\bf k}=\frac{\hbar^2 |{\bf k}|^2}{2m_{F(B)}}$ are kinetic
energies for fermions (bosons) and $\Omega$ is the volume. $U_{bf}$ is the
strength of short-range interaction between fermions and bosons, and is equal to
scattering lengths $2\pi \hbar^2 a/m_R$ after regularization \cite{Song10} (see
also Appendix A). And $m_R=m_B m_F/(m_B+m_F)$ is the reduced mass,
$\epsilon_{\bf k}^R=\hbar^2 {|\bf k|}^2/2 m_R$. This single-channel model
effectively describes near-resonance physics provided the resonance is broad
enough; i.e., the effective range is much smaller than the Fermi wavelength \cite{Pethick08}.

\begin{figure}[ht]
\includegraphics[width=\columnwidth]{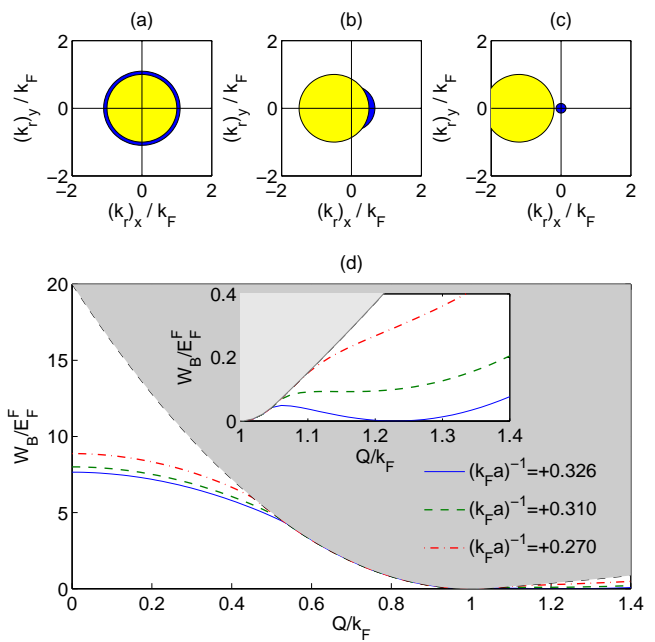}
\caption{(Color online)
(a)--(c) Schematics of the Pauli blocking effect on dimers with total momentum
$|\hbar {\bf Q}|=0$, $0.2(m_B/m_R)\hbar k_F$, $(m_B/m_R)\hbar k_F$, respectively.
Yellow (light gray) areas in the relative momentum $k_r$ space are for the blocked or occupied states
and blue (dark gray) areas stand for states available for pairing
near the threshold of the two-body continuum.
(d) Energy dispersion of excited dimers for the mass ratio $m_B/m_F=0.05$;
shown in the inset is the part of dispersion for $|{\bf Q}| > k_F$.
Shaded areas are for the two-atom excitation continuum.
At $1/k_F a=0.326$, the minimum of dispersion reaches zero energy at
a finite momentum $\hbar Q_{min}=1.23\hbar k_F$. Beyond this point, a dimer
should appear in the ground
state when a minority atom nearly resonates with a Fermi sea.
}\label{fig1}
\end{figure}

\begin{figure}[ht]
\includegraphics[width=\columnwidth]{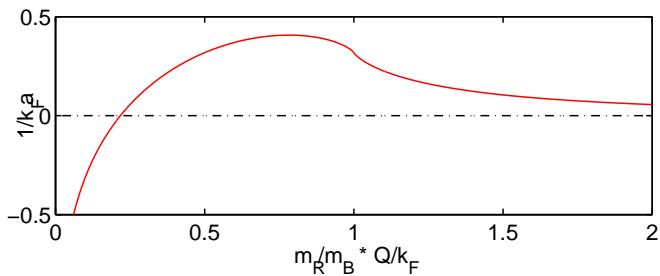}
\caption{(Color online)
For a dimer with total momentum
$\hbar Q$ to appear in the two-body excitation spectrum,
$1/(k_F a)$ has to exceed a minimum value indicated by the curve.
Dashed line is for free space without a Fermi sea.
}\label{fig2}
\end{figure}

To get insight into pairing with a nonzero total momentum $\hbar {\bf Q}$, we
examine the dispersion for a two-body bound state of Fermi-Bose atoms with an
arbitrary total momentum $\hbar {\bf Q}$ or a center-of-motion kinetic energy
$\epsilon^C_{\bf Q}=\hbar^2{\bf Q}^2/2m_T$ and $m_T=m_F+m_B$. The dispersion can
be obtained by solving the two-body problem on top of a ``frozen"
Fermi sea of majority atoms. The resultant self-consistent equation is
%\begin{eqnarray}
%W_B({\bf Q})=\epsilon_F^B+\epsilon^C_{\bf Q}+\omega_B
%\end{eqnarray}
%where
%$\epsilon_F^B=\hbar^2k_F^2/2m_B$.
%The ${\bf Q}$-dependent binding energy
%$\omega_B(<0)$ is determined by the following equation,
\begin{eqnarray}
\frac{-m_R}{2\pi\hbar^2 a}=
\frac{1}{\Omega}\sum_{\bf k}\left[
\frac{\Theta(|\frac{m_R}{m_B}{\bf Q}+{\bf k}|-k_F)}
{\epsilon_{\bf k}^R + \epsilon^C_{\bf Q}-\epsilon_F^F-W_B({\bf Q})}
-\frac{1}{\epsilon_{\bf k}^R}\right],
%\nonumber \\
\label{2bd}
\end{eqnarray}
Here $W_B({\bf Q})$ is the energy of dimers with momentum $\hbar \bf Q$, and is
measured from the Fermi energy $\epsilon^F_F=\hbar^2k^2_F/2m_F$ of majority
atoms. The unit step function $\Theta(\cdots)$ in the sum excludes occupied
states below the Fermi surface.

This equation can also be obtained by studying the scattering matrix in the
presence of a Fermi sea; the pole of the $T$ matrix on the real axis, if it exists,
corresponds to a bound state with an infinite lifetime (see Appendix A). We
find that Eq.\ref{2bd} is very similar to the well-known Cooper's solution to
two attractive electrons on the top of a Fermi sea\cite{Cooper56}. Note that the
$T$ matrix approach usually yields an additional contribution to the Cooper's
original solution due to the inclusion of hole like configurations in the
eigenvalue equation. However, since here we are dealing with a situation where
only a single minority atom resonates with majority ones and there is no
minority Fermi sea, in our case the hole like
configurations do not contribute to the binding energy.

The presence of a Fermi surface leads to an anomalous dimer excitation spectrum
when the scattering length is small and negative. Two interesting aspects of the
dispersion are worth emphasizing. First, dimers with zero momentum appear in the
form of excitations for arbitrary negative scattering lengths even when they are
small in magnitude. On the contrary, dimers with a finite momentum can exist
only when the scattering lengths exceed a critical value.
In fact, due to the Pauli blocking effect, the threshold of two-body
continuum for a total momentum $\hbar {\bf Q}$ is
\begin{eqnarray}
E_{th}({\bf Q})&=&\frac{\hbar^2 |{\bf Q}|^2}{2 m_T}+\frac{\hbar^2 q^2_{min}}{2m_R} -
\epsilon^F_F\nonumber \\
q_{min}&=& \mbox{ max}(k_F-\frac{m_R}{m_B}|{\bf Q}|, 0).
\end{eqnarray}
This is different from the case in vacuum where $E^{vac}_{th}(Q) =
\frac{\hbar^2Q^2}{2m_T}$. As a result for $Q=0$, one finds that the available
density of states $D_{2b}(E)$ for two-body scattering does not vanish when $E$
approaches $E_{th}$. When $0<|{\bf Q}| < \frac{m_B}{m_R} k_F$, one finds that $D_{2b}(E)$
always vanishes linearly as a function of energy $E-E_{th}$, i.e.
\begin{eqnarray}
D_{2b}(E)=  \frac{m^2_B k_F}{4\pi^2}\frac{E-E_{th}}{Q \left(\frac{m_B k_F}{m_R} - Q\right)}.
\end{eqnarray}
The qualitative difference between $D_{2b}(E)$ for the zero-momentum channel and for the finite-${\bf Q}$ channels
is schematically illustrated in Figs. 1(a)--(c).
Note that near the threshold, the two-atom scattering states
with zero total momentum $Q=0$ are represented by the whole shell around the Fermi surface while the scattering states with finite total momenta only correspond to a small strip around the Fermi surface.
The relatively low density of states
near $E_{th}$ for finite-$Q$ scattering puts a severe constraint
on the formation of dimers with a finite momentum.
And for a given negative scattering length, we find that dimers only appear
in the spectrum up to a maximum total momentum.
This is reflected in Fig.1(d) in which the
dimer dispersion ends up at a finite momentum and merges into the two-atom scattering continuum.
Alternatively, one can conclude that
to form a finite-momentum molecule,
the value of $(k_Fa)^{-1}$ has to exceed a minimum value as shown in Fig.2.

\begin{figure}[ht]
\includegraphics[width=\columnwidth]{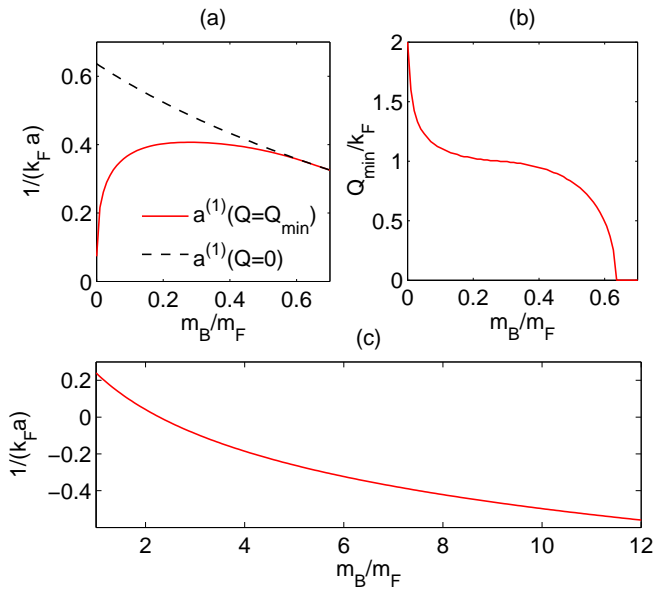}
\caption{(Color online)
(a) Critical values of $a$ at which the dimer excitation energy
$W_B({\bf Q})$ becomes zero at $Q=Q_{min}$,
the minimum of the excitation spectrum; a dimer starts
emerging in the ground state for a minority atom resonating with a Fermi sea beyond this
line.
The dashed line is obtained by setting $W_B(0)=0$ or for the scattering lengths
at which the zero-momentum dimer energy vanishes.
(c) is for larger mass ratios.
(b) $Q_{min}$, the momentum at which the dimer dispersion
touches zero for the first time when approaching the resonance from the side of negative
scattering lengths,
as a function of mass ratio $m_B/m_F$.
These plots are obtained assuming the Fermi surface is static and there are no particle-hole
fluctuations.
}
\label{fig3}
\end{figure}

Second, the dispersion minimum might locate
at a finite center-of-mass momentum $\hbar {\bf Q}$.
And this is the case either when atoms are away from resonances with
small negative scattering lengths or when Bose atoms
are very light. In the former case, it is due to the decreasing threshold
$E_{th}$ as the momentum
$Q$ increases from zero and therefore the energy of bound states below the threshold (see
Fig. 1).
In the latter situation, to form a molecule with
zero total momentum $\hbar {\bf Q}=0$ Bose atoms need to be promoted at least to
right above the Fermi surface resulting in
a high energy penalty $\epsilon_F^B$, while
to form a molecule
with total momentum near $\hbar k_F$ Bose atoms need not to be elevated.
So energetically it could be more favorable to have molecules of Fermi-Bose atoms
with a nonzero total momentum $\hbar {\bf Q}$ in this limit.

We have examined $m_{eff}$, the effective mass of extended dimers near $Q=0$,
as a function of scattering length $a$ and the mass ratio $m_B/m_F$.
At any small negative scattering length
[$-(k_Fa)^{-1} \gg 1$] or far away from resonances,
\begin{eqnarray}
\frac{1}{m_{eff}} = -\frac{m_{R}}{6m_{B}^{2}}\exp\left(-\frac{\pi}{k_{F}a}
\right)
\end{eqnarray}
and it is indeed always negative.
At scattering lengths $a^{(I)}$
where creation of a dimer with zero momentum costs no energy
or $W_B(Q=0)=0$,
we find that
\begin{eqnarray}
\frac{1}{m_{eff}}=\frac{1}{m_T}\left[1-
\frac{4m_F}{3m_B}g\left(\sqrt{\frac{m_R}{m_F}}\right)\right];
\label{effM}
\end{eqnarray}
and the dimensionless function
\begin{eqnarray}
\frac{1}{g(x)}=\left(1-x^2\right) \left[2+\frac{1-x^2}{x}\ln \frac{1+x}{1-x}\right].
\end{eqnarray}
Note that $m_{eff}$ in Eq.\ref{effM} is an indicator
of relevance or irrelevance of zero-momentum dimers when dimers start appearing in the ground state
near resonance. And
as far as $m_B/m_F > 0.7$ and the energy penalty $\epsilon_F^B$ is not too heavy,
$m_{eff}$ is
positive, although it can be much bigger than the bare mass $m_T$($=m_F + m_B$) as a
result of
dressing in the Fermi sea. In this limit, the dimer spectrum indeed crosses zero first at
$Q=0$ and a zero momentum dimer is expected to appear in the ground state.
However, when $m_B < 0.7 m_F$, the effective mass becomes negative,
implying the relevance of dimers with finite momenta in the reconstructed ground state.
Indeed we find that for small mass ratios, the minimum of the dispersion spectrum
crosses zero first at a finite momentum $\hbar Q_{min}$ when approaching resonances from the
side of
negative scattering lengths. Going further beyond this point, one would expect that
dimers with finite momenta $\hbar Q_{min}$ thus appear in the ground state for minority atoms
resonating with a Fermi sea. Details are shown in Fig. 3. In Figs. 4--5, we also present the results
for
dimers near a few broad interspecies resonances that have been studied in recent
experiments.

\begin{figure}[ht]
\includegraphics[width=\columnwidth]{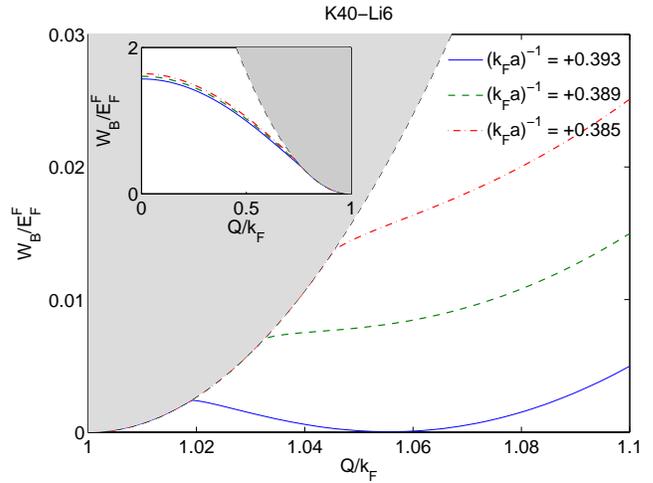}
\caption{(Color online)
Energy dispersion of dimers near an interspecies resonance of $^{6}$Li-$^{40}$K atoms.
$m_B/m_F=0.15$ for a $^{6}$Li atom nearly resonating with a Fermi sea of heavy $^{40}$K atoms.
Shaded areas are again for two-atom excitation continuum.
At $1/k_F a=0.393$,
the minimum of dispersion reaches zero energy at
a finite wave vector $Q_{min}=1.05k_F$.
}\label{fig4}
\end{figure}

\begin{figure}[ht]
\includegraphics[width=\columnwidth]{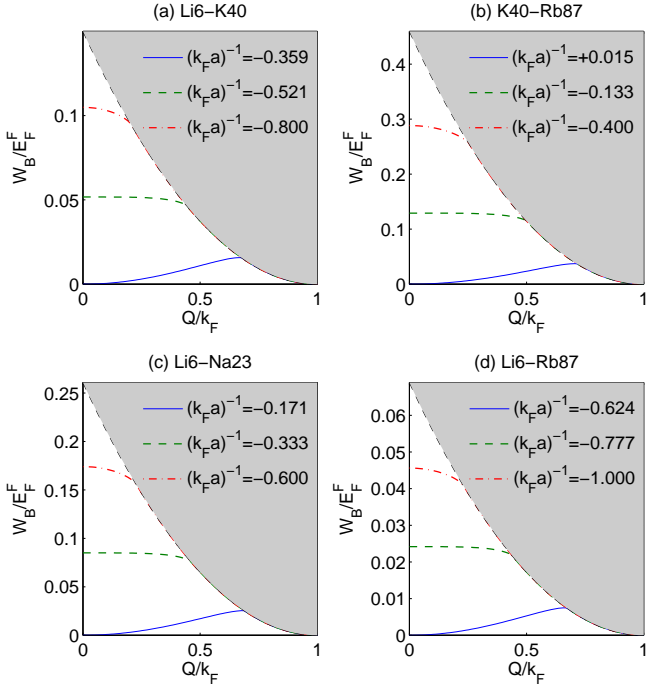}
\caption{(Color online)
Energy dispersion of dimers for different mass ratios. (a) $m_B/m_F=6.67$ for
a single $^{40}$K atom resonating with a Fermi sea of light $^{6}$Li atoms;
(b) $m_B/m_F=2.175$ or for an interspecies resonance
between $^{40}$K and $^{87}$Rb atoms;
(c) $m_B/m_F=23/6$ or $^{6}$Li-$^{23}$Na resonance;
(d) $m_B/m_F=14.83$ or $^{6}$Li-$^{87}$Rb resonance.
In (b)--(d), minority atoms are $Rb$, $Na$, and $Rb$ atoms, respectively.
Shaded areas are again for two-atom excitation continuum.
In (a)--(d), the minimum of dispersion first reaches zero energy at zero momentum.
}\label{fig5}
\end{figure}

\section{Effect of Fluctuating Particle-Hole Pairs near the Fermi Surface}

Besides the Pauli blocking effects, the energetics of bound states can be
further modified by the dynamics of Fermi seas, or virtual
particle-hole pairs near the Fermi surfaces, so the dimers are further dressed in
fluctuating particle-hole pairs.
One of the main effects of these fluctuating pairs is to renormalize
the two-body interactions near the Fermi surface and therefore to modify the binding
energetics of dimers as well as the dispersion.
The fluctuating particle-hole pairs also lead to a decay of the dimer into
either another lower energy dimer or two unbound atoms in the continuum
and we will not discuss the decay in this paper \cite{TBP}.

\begin{figure}[ht]
\includegraphics[width=\columnwidth]{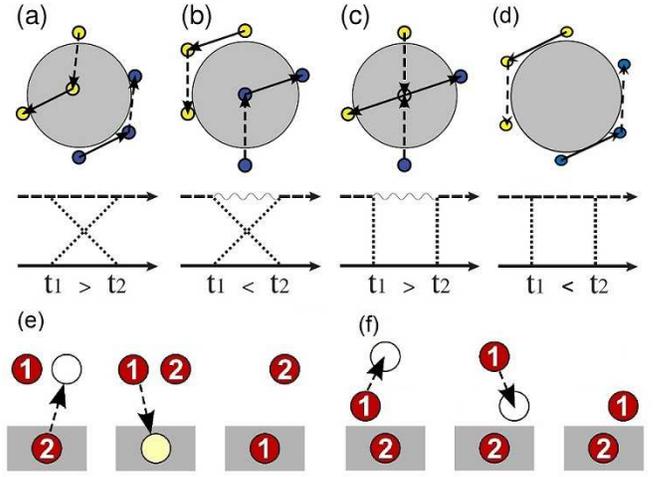}
\caption{(Color online)
An illustration of induced scattering by virtual particle-hole pairs near a Fermi surface (a),
a condensate (b), or both (c). (d) is for direct scattering without virtual particle-hole pairs.
In (a)--(d), a minority atom [blue(dark gray) dots] and majority atoms
[ yellow(light gray) ones] have an interspecies resonance with zero total momentum;
the light blue (light gray) area represents states occupied by fermions.
Also shown is a diagrammatic representation of induced scattering with the time order
illustrated explicitly. Solid line is for Fermi atoms, dashed line is for minority Bose atoms
with nonzero momentum, and wavy lines are for Bose atoms in a condensate.
In (e) and (f),
we also illustrate two distinct virtual processes
in (a)--(d): (a) and (c) involve, in addition to the
incoming Fermi atom,
a hole below the Fermi surface [as shown in (e)] while (b) or (d) does not [see (f)].
Because of the Fermi exchange statistics, the amplitude of the (e) process has an additional
negative sign
with respect to that of the direct process (d)
as well as that of the (f) process.
In the single-minority-atom limit, only A-class and the direct processes contribute;
the dashed line can also represent a minority Fermi atom in this limit.
}\label{fig6}
\end{figure}

The main effect on two-body scattering and therefore the dimer binding energy
is illustrated in Fig.\ref{fig6}, both schematically
and diagrammatically. Consider a general situation where the density for minority atoms
or Bose atoms is finite but much smaller than the density of
majority Fermi atoms \cite{Heiselberg00}.
Three classes of particle-hole pair fluctuations contribute:

(A) A majority Fermi atom is first excited from below the
Fermi surface to above by the incoming colliding minority Bose atom;
the Fermi hole is later filled by the incoming Fermi atom that simultaneously
scatters the minority Bose atom to its
final state.

(B) A minority Bose atom is excited from the condensate by the incoming colliding
majority Fermi atom;
the Bose hole is later filled
by the incoming Bose atom that simultaneously scatters the Fermi atom to its final state.

(C) A majority Fermi and a condensed Bose atom are excited with holes
left behind filled in later by the incoming majority Fermi and minority Bose
atoms.

A- and C-class processes involve exchange of virtual Fermi particle-hole pairs; i.e.,
a second majority Fermi atom has to be ejected out of the Fermi sea; they both
have an additional negative sign
with respect to the processes in class B (see Fig. \ref{fig6}) or
the direct processes without virtual particle-hole pairs.
One can also carry out a parallel analysis on a Fermi-Fermi mixture
by replacing a condensate with a Fermi sea of minority atoms.
In that case, the processes in the A-class or B-class
will have a different sign with respect to the processes in the C-class as well as
the direct scattering.
And in the limit of a single minority atom that we are going to focus on
or when the minority atom density vanishes, only the
processes in the A-class contribute.

To understand this particular energetic effect of fluctuating
particle-hole pairs, we study the $T$ matrix including
vertex corrections due to these fluctuating pairs.
The technical details of this part are summarized in Appendix B.
Scattering with the Fermi sea effectively takes momenta away from
the incoming minority atom of the dimer to create virtual particle-hole
pairs. This virtual process crucially depends on the initial and final relative momentum
between the two atoms in the dimer.
This can been illustrated via a momentum flow chart in Eq. (\ref{eq:flowchart}) for
two incoming atoms with the total momentum $\hbar {\bf Q}$ and relative momentum $\hbar {\bf k}$
scattered into two outgoing atoms with the same total momentum but
different relative momentum $\hbar {\bf k}'$ through colliding with the Fermi surface.
\begin{eqnarray}
&& f^\dagger_{\frac{m_R}{m_B}{\bf Q} -{\bf k}}
b^\dagger_{\frac{m_R}{m_F}{\bf Q} +{\bf k}}|F.S.\rangle
\rightarrow  \nonumber \\
&& f^\dagger_{\frac{m_R}{m_B}{\bf Q} -{\bf k}}
b^\dagger_{\frac{m_R}{m_F}{\bf Q} +{\bf k}-{\bf q}}
f^\dagger_{\frac{m_R}{m_B}{\bf Q} -{\bf k}'}
f_{\frac{m_R}{m_B}{\bf Q} -{\bf k}'-{\bf q}}
|F.S.\rangle
\rightarrow \nonumber \\
&&f^\dagger_{\frac{m_R}{m_B}{\bf Q} -{\bf k}'}
b^\dagger_{\frac{m_R}{m_F}{\bf Q} +{\bf k}'} |F.S.\rangle \label{eq:flowchart}
\end{eqnarray}
where $|F.S.>$ is introduced to represent the Fermi sea
of majority fermions; the virtual state energy explicitly depends on
${\bf Q}, {\bf k}, {\bf k}'$.
So with the particle-hole fluctuations, the $T$ matrix not
only depends on the total momentum $\hbar {\bf Q}$ but also
the relative momentum $\hbar {\bf k}$ and $\hbar {\bf k}'$.
Effectively, the fluctuating pairs mediate an interaction
with a finite range.

One of the ways to estimate these effects is
to introduce an effective scattering length (see Appendix B for details).
The effective scattering length $\tilde{a}$ can be expressed as
\begin{eqnarray}
\tilde{a}^{-1} &=& {a^{-1}} -k_F R + O(k^2_Fa)...,
\end{eqnarray}
Here $\tilde{a}$ differs from the free space scattering length $a$ due to the
particle-hole pair fluctuations. And $k_F R$ represents the lowest order vertex
correction induced by fluctuating particle-hole pairs near a Fermi surface (or
in condensates if a finite density of minority Bose atoms are present, or both)
as shown in Fig. \ref{fig6} (see also Appendix B). This correction is analogous to
the vertex corrections discussed for zero-momentum pairing in fermion
superfluids by Gor'kov and Melik-Barkudarov \cite{Gorkov61}. We shall denote this
correction as the GMB vertex corrections (see also \cite{Heiselberg00}) in the
following discussion.
Here we have carried out a detailed diagrammatic analysis on the GMB
correction in finite-$Q$ channels. We obtain the momentum dependence of induced
scatterings and summarize the main results below.
As a remark, we find that in the limit of heavy minority atoms (large mass
ratio $m_B/m_F$), the $R$ function for two-atom scattering with zero total
momentum is
\begin{equation}
R\rightarrow \frac{1}{\pi}\log \frac{m_B}{m_F}.
\end{equation}
We use natural logarithmic functions with base e here and throughout the paper.
This logarithmic divergence in the large-mass-ratio limit can be attributed to
the very heavy dressing of a zero-momentum dimer in virtual particle-hole pairs
near the Fermi surface. One can also demonstrate that this has the same origin as
the Anderson's infrared catastrophe in a quantum impurity problem \cite{Anderson67}.

\begin{figure}[ht]
\includegraphics[width=\columnwidth]{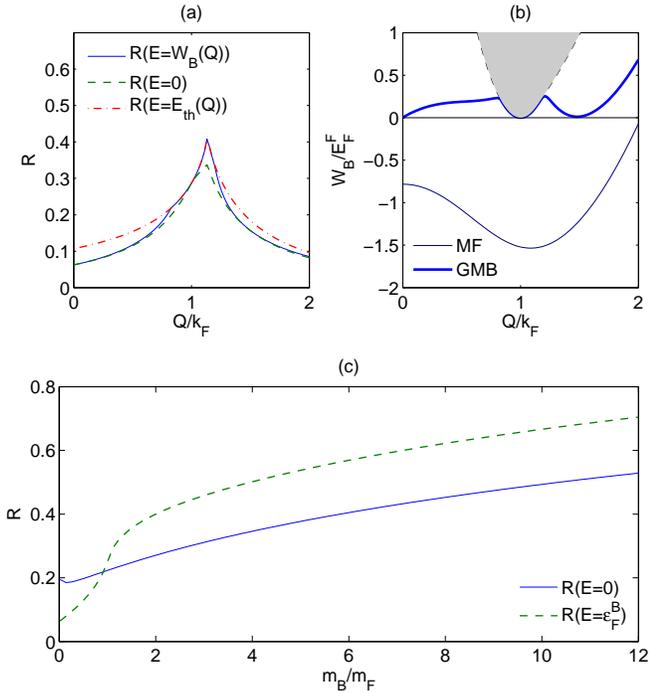}
\caption{(Color online)
(a) $R$, the vertex correction
as a function of total momentum $\hbar Q$ for mass ratio $m_B/m_F=0.135$. Here the
parameter
$R$ is defined as $1/k\tilde{a}=1/k_Fa -R$.
Different plots are obtained with different choices of energy $E$;
the solid line is obtained using a self-consistent method where $E=W_B(Q)$.
(See more discussions in Appendix B.)
(b) Dispersion of dimers with GMB
and without MF particle-hole fluctuations or formally Gorkov-Melik-Barkudarov
(GMB)
vertex corrections for $m_B/m_F=0.135$
and $(k_F a)^{-1}=0.62$.
(c) $R$ for the $Q=0$ scattering channel as a function of
mass ratio $m_B/m_F$.
}\label{fig7}
\end{figure}

We have numerically found that for scatterings near the threshold of
two-particle continuum, the exchange processes in the A-class described above
always induce an effective {\em repulsive} interaction. So the fluctuating
particle-hole pairs in this case effectively {\em screen} or {\em reduce} the
interspecies attractive interactions in all channels of momenta $\hbar \bf Q$ (see
Fig.\ref{fig7}).
Furthermore, the {\em magnitude} of the corrections to the
effective two-atom scattering strongly depends on the total momentum $\hbar \bf
Q$ of the scattering atoms.
This can be demonstrated by explicitly examining two
cases: $\hbar Q=0$ and $\hbar Q =\hbar k_F$ in the limit of {\em light} minority atoms,
i.e., when the energy of virtual states is dominated by the minority atom.
For a pair of minority and majority atoms in the $\hbar Q=0$ channel, or with momentum $(\hbar {\bf
k},-\hbar {\bf k})$, it requires that the minority atom momentum be close to
$\hbar k_F$ because of the Pauli blocking effect.
Consider such an incoming minority atom with momentum $\hbar {\bf k}$
colliding with the Fermi surface popping up a pair of particle-hole excitation.
The intermediate virtual state with total energy $\epsilon_{v}$, and total momentum $\hbar {\bf k}$
consists of an outgoing minority atom and a particle-hole pair of majority atoms.
This energy $\epsilon_{v}$ can be either larger or smaller than the energy of the incoming
minority atom $\epsilon^B_{\bf k}$.
The contribution to the effective interactions from a virtual state of energy $\epsilon_{v}$
is inversely proportional to $\epsilon^B_{\bf k} -\epsilon_{v}$ and so the contributions from
different virtual states would have different signs depending on whether
$\epsilon^B_{\bf k}$ is larger or smaller than $\epsilon_{v}$, thus leading to a
destructive interference between different configurations.
Now let us turn to the scattering channel with the total momentum near
$\hbar |{\bf Q}|=\hbar k_F$, i.e. a minority atom with $\hbar {\bf k}=0$
interacting with a majority atom near the Fermi surface with Fermi momentum $\hbar k_F$.
For a minority atom with $\epsilon^B_{\bf k}=0$, {\em all} intermediate virtual states would have
energies larger than the initial one and all should contribute to the effective
interaction with the same sign, leading to a constructive interference.
This results in a maximum value of the correction to two-body interactions; i.e.,
the correction reaches a peak value when ${\hbar \bf k}$ is near zero or $\hbar
|\bf Q|$ is near $\hbar k_F$.

A more elaborated examination along this line can be carried out in Appendix B.
There we show that for all virtual states involving a hole like excitation of
momentum $\hbar \bf l$, the most dominating contribution to the overall vertex
correction is always from the state with $\hbar {\bf l} \simeq \hbar k_F {\bf
Q}/|{\bf Q}|$ so that the energy difference between the virtual state and the
initial state is minimum.
We can further introduce an effective {\em group} velocity for virtual states
involving a hole like excitation with momentum as
${\hbar} {\bf l}$, $\partial \epsilon_v({\bf l})/\partial {\bf l}$.
For the scattering near the threshold of the two-body continuum, this velocity turns out to
be proportional to $m_R/m_B Q-k_F$.
In the limit of light minority atoms, i.e., the reduced mass $m_R$ approaches $m_B$,
the dominating virtual states appear softest near $\hbar Q=\hbar k_F$, consistent with the above qualitative argument
based on interference effects.
In general, formation of bound states near $\hbar Q=(m_B/m_R) \hbar k_F$ is
disfavored by the fluctuating particle-hole pairs.
For this reason, the vertex correction to two-body interactions
due to the creation of particle-hole pairs is relatively small near
small and large total momentum, and peaks near $\hbar Q= \hbar k_F$
in the limit of light minority atoms.
This is confirmed in our numerical evaluations of $R$.
So fluctuating particle-hole pairs favor the formation of
dressed bound states with $\hbar Q$ away from $\hbar k_F$,
over the dimers with $\hbar Q\simeq \hbar k_F$.
Note that although we arrive at this conclusion by considering only the lowest
order corrections, we conjecture that this is generically true as far as the
effective group velocity for the hole like excitations has a minimum near
$\hbar Q= \hbar k_F$. The position of the peak in Fig.\ref{fig7} in our opinion correlates
with the threshold minimum where the screening effect
is usually the strongest.

The problem of two-atom scattering near $\hbar Q = \hbar k_F$ is closely related to the spin polaron in
cold gases that has been discussed quite extensively in the recent
literature\cite{Chevy06,Pilati08,Combescot07,
Prokof'ev08,Punk09,Mora09,Bruun08,Schirotzek09}.
In our calculation, we also see a transition from a polaron to dimer when
approaching the resonance from the negative-scattering-length side.
We refer the readers to those publications for more elaborated discussions on the spin
polaron.
However, we should also emphasize that although the physics of the spin polaron
itself is an interesting subject, to understand the nature of the ground state of a single minority
atom; or the general pairing correlations in quantum mixtures near resonances,
it is necessary to study the whole dimer spectrum.
This is because as we have seen before, the minimum of the dispersion can occur
at any momentum depending on the mass ratio $\frac{m_B}{m_F}$.
The minimum location $\hbar Q_{min}$ generally differs from $\hbar Q=\hbar k_F$ or $\hbar Q=0$.
Only for a certain range of mass ratios, scatterings near $\hbar Q=0$ or $\hbar Q=k_F$ set the
overall instability of a unpaired minority atom.

\begin{figure}[ht]
\includegraphics[width=\columnwidth]{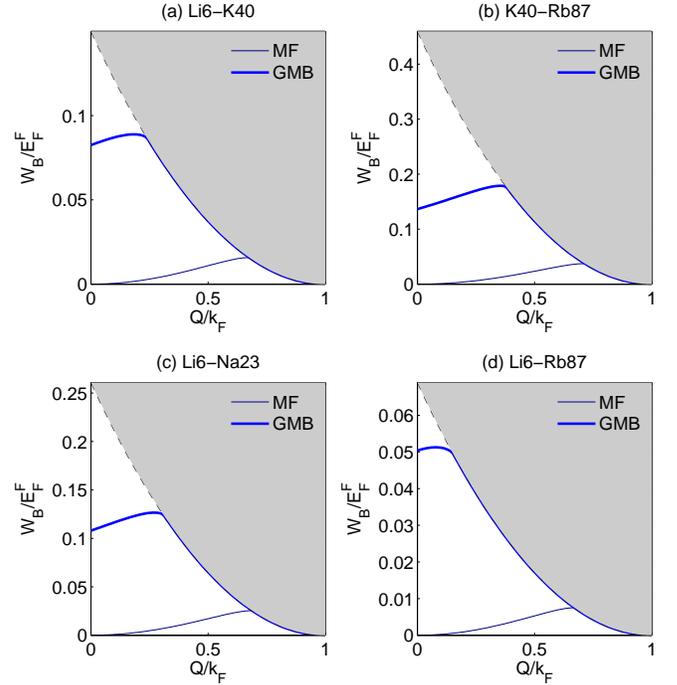}
\caption{(Color online)
The vertex corrections to the dimer spectrum (denoted as GMB).
Thin solid lines are the reference spectrum
without the vertex correction, or a mean field (MF) result.
(a) $m_B/m_F=6.67$ for
a single $^{40}$K atom resonating with a Fermi sea of light $^{6}$Li atoms;
(b)$m_B/m_F=2.175$ or for an interspecies resonance
of $^{40}$K-$^{87}$Rb atoms;
(c) $m_B/m_F=23/6$ or $^{6}$Li-$^{23}$Na resonance;
(d)$m_B/m_F=14.83$ or $^{6}$Li-$^{87}$Rb resonance.
In (b)--(d), minority atoms are Rb, Na, and Rb atoms respectively.
Shaded areas are again for two-atom excitation continuum.
}\label{fig8}
\end{figure}

\section{Implications}

The dispersion obtained for an individual minority atom having interspecies
resonance with majority Fermi atoms that form a Fermi sea can be applied to
identify the instability point\cite{instability} for an extremely imbalanced quantum Fermi-Bose or
Fermi-Fermi mixture beyond which the weakly interacting mixtures become
unstable. Below we assume the minority species can be either Fermi or Bose atoms
with densities much lower than that of majority atoms.

The instability point corresponds to the scattering length at which the
dispersion of anomalous dimers touches zero or the energy cost of creating a
dimer becomes zero. Since the minimum in the dimer dispersion can be either at
${\bf Q}=0$ or ${\bf Q} \neq 0$, it is implied that the instability might be
driven by formation of dimers with either ${\bf Q}=0$ or a finite ${\bf Q}$.

We summarize the results of the dispersion for different mixtures in Fig. 8,
and the instability for different mass ratios in Fig. 9
Since the fluctuating particle-hole pairs near the Fermi surface reduce the
binding energy of dimers, the formation of a dimer in the ground state for a
minority atom and a Fermi sea of majority atoms occurs at a higher critical
value of $1/k_Fa$ when crossing the resonance from the side of negative
scattering lengths. The second effect of fluctuations here appears to
destabilize dimers with finite momenta near $k_F$, because of the peak structure
in the vertex corrections. This is qualitatively consistent with the results in
Fig. 9(b): The vertex corrections suppress the area of the region where $Q_{min}$ is
finite.

For the case of equal-mass interspecies resonance, the mean field calculation
with a static Fermi surface yields a critical value, $(k_F a)^{-1}=0.24$;
fluctuations of particle-hole pairs correct the mean field instability point
leading to a modified critical value $(k_F a)^{-1}=0.34$. Compared to $(k_Fa)^{-1}=0.88$,
the critical value obtained via the diagrammatic quantum Monte
Carlo \cite{Prokof'ev08} that has so far been done only for the equal-mass case,
our approximation yields a qualitative correct account of the main effects of
fluctuations but quantitatively underestimates the effects of fluctuations.
And in our diagrammatic approach, for mass ratios ($m_B/m_F$) less than $0.135$
a dimer with finite $\bf Q$ appears in the ground state for a minority atom
resonating with a Fermi sea of majority atoms at a critical value of scattering
length $a$. This is close to $0.15$, the numerical result obtained in a
truncated particle-hole pair subspace \cite{Huse10}. However for this critical
mass, the critical scattering length at which a dimer starts forming in the
ground state is predicted as $(k_Fa)^{-1}=0.62$ in our approach while in
Ref. \cite{Huse10}, $(k_Fa)^{-1}$ is close to a larger value of $1.7$.

\begin{figure}[ht]
\includegraphics[width=\columnwidth]{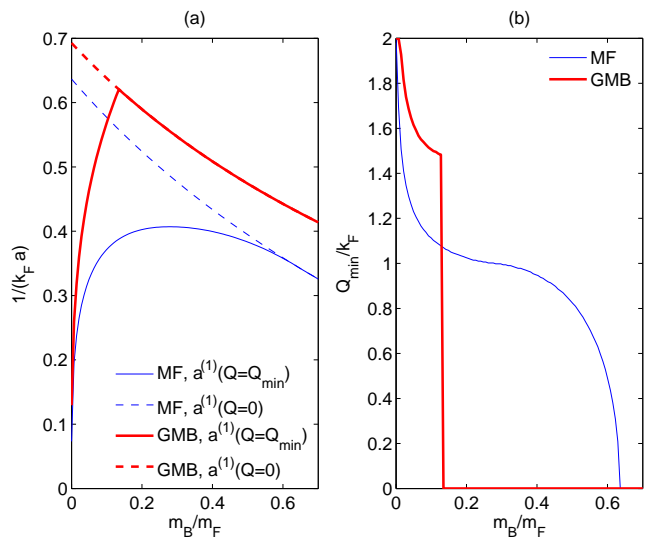}
\caption{(Color online)
The estimated effect of particle-hole fluctuations(curve labeled as GMB)
on the dimer formation;
the result obtained in the static Fermi surface approximation is denoted
as MF.
(a) Critical line beyond which a dimer starts to appear in the ground state for a minority
atom resonating with a Fermi sea of atoms;
the dashed lines are obtained when only taking
into account the zero-momentum dimers, i.e.
set by scattering lengths at which $W_B({\bf Q}=0)$ vanishes.
(b) The momentum of a dimer that appears in the ground state right
above the critical line.
}\label{fig9}
\end{figure}

It remains to be understood
what happens beyond the instability line
identified in Fig.\ref{fig9} and what is the nature of the
transitions if there are any \cite{PT}.
For Fermi-Bose mixtures, an earlier attempt was made to understand dimer correlations beyond
a critical line \cite{Watanabe08} that was obtained when the scatterings in finite-$\bf Q$ channels were not included.
However, the appearance of finite-density fermion dimers in the ground state beyond the critical line was
not properly taken into account in analyses.
This complication and more importantly
potential dimer-dimer and atom-dimer interactions
about which we know very little make a thorough study in this limit extremely challenging.
The second issue is how dimers in
a finite-$|{\bf Q}|$ channel compete against
other structures with higher order correlations.
For instance, it was also
proposed that for two-dimensional Fermi-Fermi mixtures with extremely small mass ratios
less than $0.01$ \cite{Petrov07}, a crystal structure can further develop at exponentially
low temperatures.
This question perhaps can be best addressed
by numerically taking into account various fluctuations.
On the other hand, we remark that LOFF states have been suggested in Fermi gases near
resonances.
For Fermi-Fermi mixtures with equal masses, Bulgac {\it et al.} suggested that the ground
state
can be a pairing state that breaks
translational symmetries \cite{Bulgac08}.
Recently, it was also argued that LOFF states appear on the molecular
side of resonances in almost fully polarized Fermi gases but only for nonequal masses;
in addition, a trimer phase was also proposed \cite{Huse10}.
It remains to be clarified the connections between different results obtained in
Ref. \cite{Petrov07}, Ref. \cite{Bulgac08} and Ref. \cite{Huse10}.
In Ref. \cite{Song10}, the authors first investigated the
anomalous dispersion of molecules in Fermi-Bose mixtures and estimated the
effect of fluctuating particle-hole pairs on the formation of dimers.
The authors then proposed a first-order phase transition between the noninteracting
ground state and
a fully paired state analogous to the BCS one.
Dukelsky {\it et al.} \cite{Dukelsky10} correctly pointed out that a string factor was missing in the energetic analysis
of the first-order phase transition.
The correct form of the energy function should be
\begin{eqnarray}
E = &&\sum_{\bf k} \left(|v_{\bf k}|^2 + |\eta_{\bf k}|^2\right)\epsilon^F_{\bf k}
+ \sum_{\bf k} |v_{\bf k}|^2\epsilon^B_{\bf k} \nonumber \\
&&+ \frac{U_{bf}}{\Omega} \sum_{\bf k k'}
u^*_{\bf k'} v^*_{\bf k'} u_{\bf k} v_{\bf k} \times \nonumber \\&&
\prod_{ \bf k'' \in (k,k')}
\left(|u_{\bf k''}|^2-|v_{\bf k''}|^2 -|\eta_{\bf k''}|^2\right).
\end{eqnarray}
Here the string product is carried over states between ${\bf k}$ and ${\bf k'}$ in
the trial wave function given in Eq.(5) in Ref. \cite{Song10}.
In the previous calculations, the string product was indeed overlooked and
its effect was not properly taken into account.
The string factor modifies the interaction energy via inducing a rapidly alternating sign when
the momentum varies,
and more importantly unless the distribution function is strictly the Fermi-Dirac function, the
string product should be equal to zero in the thermodynamic limit where there are an
infinite number of states between the two open ends of the string.
For this reason, the energy of the proposed pairing states near the first-order phase
transition was erroneously underestimated and within this mean field approximation with
the string product properly taken into account,
one is no longer able to show that
there is a first-order phase transition to the proposed pairing state as argued in the previous
Letter \cite{Song10}.
In Ref. \cite{Dukelsky10},
the Richardson solution has also been further utilized to
argue the nonexistence of collective states below the energy of zero-momentum molecules.
It was pointed out that Fermi-Bose molecules are noninteracting
in the reduced pairing model employed there; in that reduced Hamiltonian
that differs from the full scattering Hamiltonian [see Eq.(1)] introduced here,
scatterings in all finite-$Q$ sectors are completely suppressed.
So the result of the nonexistence of lower collective modes,
although it is a feature of the reduced Hamiltonian, might not be a property
of cold atoms near Feshbach resonances. In fact, one can show that the second-order
commutator-anticommutator for composite molecules does not vanish when the full
Hamiltonian in Eq.(1) rather than the reduced pairing model is
considered \cite{commutator}. Two-body scattering with a finite total momentum $\hbar Q$
that was neglected in the reduced pairing Hamiltonian induces significant dimer-atom
or dimer-dimer scatterings as well as scatterings between a dimer and a Fermi
sea. In Fermi-Fermi mixtures, the finite-$Q$ scattering also plays a critical
role in dimer-atom or dimer-dimer scattering near resonances. The dimer or
molecule dynamics near resonance can be understood when these scatterings are
properly taken into account. For this reason, the issue of low-lying collective
modes of the molecules remains an open question and can be addressed when
the physics beyond the Richardson-solution is considered.

In summary, to understand the physics beyond the instability line, it is
essential to include the two-body scattering with a finite total momentum $\hbar
\bf Q$ when locating  the critical scattering length for the molecule dispersion can be quite
anomalous and its
mass near $Q=0$ can be negative either when far away from resonance or when the
minority atom is very light. In the latter case, one can show that Fermi-Bose
molecules with finite momenta should first appear in the ground state because
$Q=0$ molecules are energetically unfavorable; the instability of the
noninteracting ground state is therefore driven by  scatterings in finite-$Q$ sectors
rather than two-body scattering processes in the reduced pairing Hamiltonian,
which excludes two-atom scattering with finite $Q$.

\section{Conclusion}

In conclusion, we have investigated dressed dimers in quantum mixtures and
carried out thorough studies of energies of dimers in finite-${\bf Q}$ channels.
Our calculation suggests that in the limit of light bosons, dimers with a finite
momentum $\hbar {\bf Q}$ are most relevant for the ground state for a minority
atom resonating with a Fermi sea of majority atoms. There are two open issues
that need to be understood in the future, most likely beyond the framework of
the approach here. One is the role of three-body correlations in quantum
mixtures. Following a general argument by Efimov, when $m_B/m_F$ mass ratios are
smaller than $0.145$, trimers involving two heavy majority Fermi atoms should
form in vacuum \cite{Efimov72}. However, three-body dimer-atom resonances induced
by trimers generally are much narrower than two-body resonances and
contributions from Efimov physics to the two-body effective interactions can be
smaller than the universal corrections that originate from collective
excitations \cite{Gorkov61} which we have taken into account in this article.
This was previously noticed in the studies of Fermi-Fermi
mixtures \cite{Baranov08}. In this limit, one should expect that trimers have
relatively weak effects on the physics in two-body channels and that dominating
effects in the dimmer channel discussed above can
survive these higher order corrections.

A related and perhaps more important question is under
which conditions three-body correlations are dominating
and an additional trimer channel
has to be added to the discussion on many-body physics.
How trimers mediate additional strong scattering
between dimers or dimer and atoms and
lead to nontrivial higher order many-body correlations
represents one of the most challenging issues we face now and
remains to be studied in the future.
The answer to this question should depend on
short-distance behaviors of Efimov potentials and would therefore be nonuniversal.
Our results about dimers should be relevant whenever these nonuniversal higher order
correlations are insignificant, and the fraction of atoms forming trimers
is small.

We acknowledge valuable discussions with Mikhail Baranov, Georg Bruun,
Eugene Demler, Chris Greene, Rudi Grimm, Carlos Lobo, David MacNeill,
Mohammad Mashayekhi, Subir Sachdev, Klaus Sengstock, Dam Thanh Son, Joseph Thywissen,
Vladan Vuletic, Matteo Zaccanti, Peter Zoller, and Martin Zwierlein.
One of the authors (F.Z.) would like to thank
the physics department, at Harvard University and the
Center for Ultracold Atoms, at MIT and Harvard for their hospitality
during his visit in the winter and spring of 2011, when this article was finalized.
This work is in part supported by the Canadian Institute for Advanced
Research, the Izaak Walton Killam Fund for Advanced Studies, NSERC (Canada) and CUA at MIT
and Harvard, and the Austrian Science Fund FWF FOCUS.

\begin{widetext}

\appendix

\section{$T$ matrix in the static-Fermi-sea approximation and
the molecule's effective mass}

\begin{figure}
\includegraphics[width=\columnwidth]{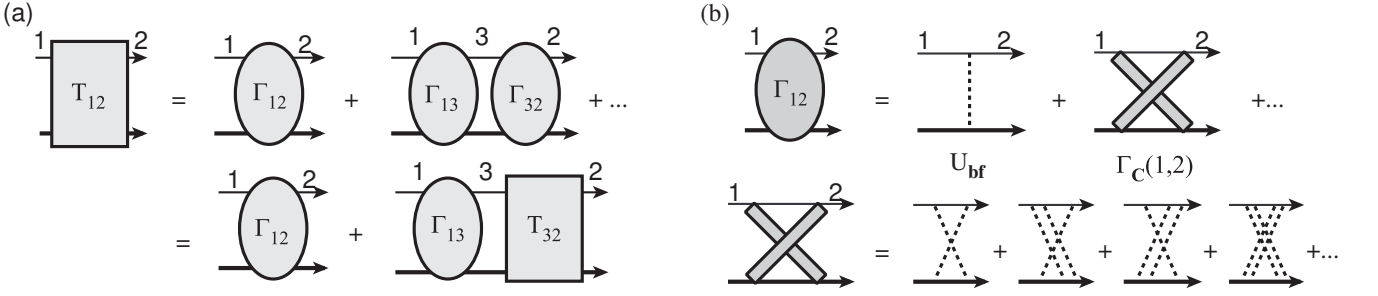}

\caption{\label{fig:Diagrams} Feynman diagrams for the $T$ matrix calculation,
thick (thin) solid lines are for majority (minority) atoms, dashed
lines are for the bare interaction $U_{bf}$, and $\Gamma(\cdots)$ is
the irreducible two-body interaction vertex. (a) The ladder summation
and diagrammatic representation of the Bethe-Salpeter equation in Eq. (\ref{eq:bse}).
(b) $\Gamma$ with GMB corrections. Here the interaction lines within
the $\Gamma_{C}(1,2)$ are renormalized interactions, which can be
expanded in series of ladder diagrams using bare interaction $U_{bf}$.}
\end{figure}

To determine the molecular excitation energy in a Fermi sea of majority
atoms, we look at the $T$ matrix for the minority-majority scattering.
Diagrammatically it is the skeleton diagram (without the external
propagators) of the following time-ordered two-particle propagator:
\begin{eqnarray}
\langle F.S.|T c(t,\mathbf{k}_{2})b(t,{\bf Q-k_{2}})b^{\dagger}(0,{\bf Q}-
\mathbf{k}_{1})c^{\dagger}(0,{\bf k_{1}})|F.S.\rangle
\end{eqnarray}
where $|F.S.\rangle$ stands for the Fermi sea of majority atoms. An
isolated pole of $T$ in the energy space represents a dimer excitation.
The pole could be either on the real axis or in the lower complex
half plane; in the latter case the dimer has a finite lifetime.
Note that the above time-ordered propagator is zero when $t<0$, because
one cannot create a hole excitation of minority atoms.

In the following we use the ladder approximation to calculate $T$
as shown in Fig.\ref{fig:Diagrams}. The rung of each ladder
is the $\Gamma$ function or the irreducible two-body interaction vertex,
and the side rails are Green's functions of majority and minority atoms.
One can write down the Bethe-Salpeter equation to include
all the ladder-like diagrams as the following:
\begin{eqnarray}
T(1,2) & \equiv & T(\omega_{1},{\bf k}_{1},E-\omega_{1},Q-{\bf k}_{1};\omega_{2},{\bf
k}_{2},E-\omega_2, Q-{\bf k}_{2})\\
H(3) & \equiv & iG_F(\omega_{3},{\bf k}_{3})G_{B}(E-\omega_{3},Q-{\bf k}_{3})\\
T\left(1,2\right) & = &
\Gamma(1,2)+\sum_{3}\Gamma(1,3)H(3)T(3,2)\label{eq:bse}\end{eqnarray}
where we have introduced a shorthand notation $(1,2)$ to denote the
incoming and outing states.
Here $G_F$ ($G_B$) are bare Green's functions for majority(minority) atoms defined as,
$G^{-1}_B(\omega,{\bf k}) \equiv \omega-\epsilon^B_{\bf k} + i0^+$
and $G^{-1}_F(\omega,{\bf k}) \equiv \omega-(\epsilon^F_{\bf k}-\epsilon^F_F)(1 - i0^+)$.
In the leading order approximation, we
have $\Gamma(1,2)=U_{bf}$, and one can further integrate over $\omega_{3}.$
The result is
\begin{eqnarray}
\frac{1}{T(1,2)} & = & \frac{1}{U_{bf}}+\frac{1}{\Omega}\sum_{\bf k}\frac{1-n_{F}({\bf
k})}{\epsilon_{\bf k}^{F}+\epsilon_{\bf Q-k}^{B}-\epsilon_{F}^{F}-E-i0^{+}}\label{eq:invT}
\end{eqnarray}
The energy of a dimer is determined from the condition that $T$ becomes divergent when
$E$ approaches $W_B(Q)$:
\begin{eqnarray}
-\frac{m_{R}}{2\pi\hbar^2 a} & = & \frac{1}{\Omega}\sum_{\bf k}\left[\frac{1-n_{F}({\bf
k})}{\epsilon_{\bf k}^{F}+\epsilon_{\bf Q-k}^{B}-\epsilon_{F}^{F}-W_B(Q)}-
\frac{1}{\epsilon_{k}^{R}}\right]\label{eq:invT-re}\end{eqnarray}
The last term in Eq. (\ref{eq:invT-re}) was introduced to regularize the ultraviolet divergence.

The integral in Eq. (\ref{eq:invT-re}) can be evaluated analytically for an arbitrary value of
$W_B(Q)$. And the result for the bound-state solution when $W_B(Q) < E_{th}(Q)$ is
\begin{eqnarray}
1-\frac{\pi}{k_{F}a} & = & -\frac{1-2b^{2}+c}{4b}\log\left(\frac{1+2b+c}{1-2b+c}\right) +
\sqrt{b^{2}-c}\left[\pi I +  {\rm arctanh}\frac{1-b}{\sqrt{b^2-c}} + {\rm arctanh}\frac{1+b}{\sqrt{b^2-c}}\right]
\\
b & \equiv & \frac{1}{\epsilon_{F}^{R}}\frac{\hbar^2 k_{F}Q}{2m_{B}},\textrm{ }
c\equiv\frac{1}{\epsilon_{F}^{R}}\left(\frac{\hbar^2Q^{2}}{2m_{B}}-\epsilon_{F}^{F}-
W_B\right)\nonumber \end{eqnarray}
From the above expression one can find the effective mass of the molecule $m_{eff}$
defined as $W_B(Q)-W_B(0)\simeq\frac{\hbar^2 Q^{2}}{2m_{eff}}$,
\begin{eqnarray}
\frac{1}{m_{eff}} & = & \frac{1}{m_{B}+m_{F}}-\frac{2m_{R}}{3m_{B}^{2}}\frac{1}{1-
y^{2}}\frac{1}{1+\frac{1-y^{2}}{2y}\log\frac{1+y}{1-y}}\nonumber \\
y &\equiv& \sqrt{\frac{W_B(Q=0)+\epsilon^F_F}{\epsilon^R_F}}
\end{eqnarray}
The effective mass at the instability point given in Eq. (\ref{effM}) is obtained by setting $W_B(Q=0)=0$
or $y=x=\sqrt{\frac{m_{R}}{m_{B}}}$. One can also find out that the effective mass is always
negative
in the limit where $\frac{1}{k_{F}a}\rightarrow-\infty$ by studying the asymptotics when
$y\rightarrow 1 - 2\exp\left({\frac{\pi}{k_Fa}}\right)$:
\begin{equation}
\frac{1}{m_{eff}}\rightarrow -\frac{m_{R}}{6m_{B}^{2}}\exp\left(-
\frac{\pi}{k_{F}a}\right)
\end{equation}

\section{Effective scattering length $\tilde{a}$ From $T$ matrix Method}

In this section we use the $T$ matrix method to study the effects of
fluctuating particle-hole pairs near Fermi surface. Our analysis shows
that in the limit of small $k_{F}a$, the effects due to particle-hole
fluctuations can be captured by introducing an effective scattering length
$\tilde{a}$ as $\tilde{a}^{-1}=a^{-1}-k_{F}R + k_{F}O(k_{F}a)$, where the function
$R$ represents the lowest order vertex correction.
Similar quantities such as effective interactions have been introduced before \cite{Heiselberg00},
to address effects of particle-hole fluctuations in zero-momentum pairing physics.
The $T$ matrix method used here provides a unified description for dimers
of different momenta and energies, and can be further extended to near resonances \cite{TBP}.
In the following we first show how to obtain Eq. (\ref{eq:bse-gmb})
for dimer's energy by examining the pole structure of the $T$ matrix.
Then we show how to obtain Eq. (\ref{eq:gap-equation-gmb}) where an
effective scattering length  emerges.

We start from Fig.\ref{fig:Diagrams}(b), where we include the leading
order term $\Gamma_{C}(1,2)$ as
\begin{eqnarray}
\Gamma(1,2) & \simeq & U_{bf}+\Gamma_{C}(1,2)\\
\Gamma_{C}(1,2) & \simeq & i\left(\frac{2\pi\hbar^2 a}{m_{R}}\right)^{2}
\frac{1}{\Omega}
\sum_{\bf l}\int \frac{d\omega}{2\pi}G_{F}(\omega,{\bf l})G_{B}(\omega-\omega_{1}+E-
\omega_{2},{\bf Q}-{\bf k}_{1}-{\bf k}_{2}+{\bf l})\nonumber \\
&=&\left(\frac{2\pi\hbar^2 a}{m_{R}}\right)^{2} \frac{1}{\Omega} \sum_{\bf l}
\frac{n_F({\bf l})}{E-\omega_1-\omega_2-\epsilon^F_{F} +\epsilon^F_{\bf l}- \epsilon^B_{\bf
Q +l -k_1 -k_2} +i0^+}
\label{eq:gmb-def}
\end{eqnarray}
Here $\Gamma_{C}(1,2)$ corresponds to exactly the type-A virtual process
discussed in the main text. Note that in $\Gamma_{C}(1,2)$ we replace
the renormalized interaction vertex (which is a sum of diagrams with
bare interactions) for particles or holes near the Fermi-surface by two-body
scattering amplitude in vacuum, i.e., $\frac{2\pi\hbar^2 a}{m_{R}}$. This
is justified in the limit of small $k_{F}a$; such a replacement
will not change function $R$, but only leads to inaccuracies in higher
order terms. With Eq. (\ref{eq:bse}) one can write down the corresponding
Bethe-Salpeter equation:

\begin{eqnarray}
T\left(1,2\right) & = &
\left[U_{bf}+\Gamma_{C}(1,2)\right]+\sum_{3}\left[U_{bf}+\Gamma_{C}(1,3)\right]H(3)T(3,2
)\end{eqnarray}
The above equation defines an infinite series implicitly in terms
of $U_{bf}$ and $\Gamma_{C}$. It is not hard to regroup terms in the
series according to the powers of $U_{bf}$ and sum them up in the following way as
\begin{eqnarray}
T(1,2) & = & \frac{\phi_{M}(1)\phi_{M}(2)}{U_{bf}^{-1}-
\left[\sum_{3}H(3)+\sum_{34}H(3)\Gamma_{C}(3,4)H(4)+\sum_{345}H(3)\Gamma_{C}(3,4)H
(4)\Gamma_{C}(4,5)H(5)+\cdots\right]}\nonumber \\
&&+\left\{\Gamma_{C}(1,2)+\sum_{3}\Gamma_{C}(1,3)H(3)\Gamma_{C}(3,2)+\sum_{34}\Gamma_{C}(1,3)H(3)\Gamma_{C}(3,4)H(4)\Gamma_{C}(4,2)+\cdots\right\} \\
\phi_{M}(1) & \equiv &
1+\sum_{3}\Gamma_{C}(1,3)H(3)+\sum_{34}\Gamma_{C}(1,3)H(3)\Gamma_{C}(3,4)+\sum_{
345}\Gamma_{C}(1,3)H(3)\Gamma_{C}(3,4)H(4)\Gamma_{C}(4,5)+\cdots\end{eqnarray}
We can assume that the series in the curly brackets and $\phi_{M}$
are convergent, based on the estimation that each term is at most
of order $(k_{F}a)$ compared the preceding term. Then the pole in
the $T$ matrix is completely determined by the denominator, and the dimer's
energy $E$ should satisfy the following equation:
\begin{eqnarray}
\frac{1}{U_{bf}} & = & \sum_{3}H(3)+\sum_{34}H(3)\Gamma_{C}(3,4)H(4)+\cdots\\
&\simeq& \frac{1}{\Omega}\sum_{\bf k}
\frac{1-n_{F}({\bf k})}{\epsilon_{\bf k}^{F}+\epsilon_{\bf Q-k}^{B}-\epsilon_{F}^{F}-E}
+ \frac{1}{\Omega^{3}}\sum_{\bf k,p}\frac{1-n_{F}({\bf k})}{\epsilon_{\bf
k}^{F}+\epsilon_{\bf Q-k}^{B}-\epsilon_{F}^{F}-E}\frac{1-n_{F}({\bf p})}{\epsilon_{\bf
p}^{F}+\epsilon_{\bf Q-p}^{B}-\epsilon_{F}^{F}-E}\nonumber \\
 & & \times \left(\frac{2\pi \hbar^2 a}{m_{R}}\right)^{2}\sum_{\bf l} \frac{
(-1) n_{F}({\bf l})}
{\epsilon_{\bf k}^{F}+\epsilon_{\bf p}^{F}-\epsilon_{\bf l}^{F}-\epsilon_{F}^{F}+\epsilon_{\bf
Q+l-k-p}^{B}-E-i0^+}\label{eq:bse-gmb}\end{eqnarray}
Here in Eq. (\ref{eq:bse-gmb}) we have only kept leading order terms,
and we have further integrated over frequencies $\omega_{3}$ and
$\omega_{4}$.

Compared to Eq. (\ref{eq:invT}) for the static Fermi sea, the above equation
is very similar except for an additional triple-momentum integral
due to particle-hole fluctuations. We start to analyze this triple
integral by examining the range of various denominators. We first notice
that $\epsilon_{\bf k}^{F}+\epsilon_{\bf p}^{F}-\epsilon_{\bf l}^{F}-
\epsilon_{F}^{F}+\epsilon_{\bf Q+l-k-p}^{B}-E$
represents the energy difference between the dimer and virtual processes,
and hence can be either positive or negative depending on the values
of $\bf k,p,l$ (one exception is at $E=0$, the denominator is always
positive for all possible $k,p,l$). On the contrary, the other two
denominators, i.e., $\epsilon_{\bf k}^{F}+\epsilon_{
\bf Q-k}^{B}-\epsilon_{F}^{F}-E$
and $\epsilon_{\bf p}^{F}+\epsilon_{\bf Q-p}^{B}-\epsilon_{F}^{F}-E$, are
always positive for all possible values of $\bf k,p$, which is guaranteed
by the definition of bounded dimers; furthermore in the limit $k_{F}a\rightarrow0^{-}$,
these two denominators can be very close to zero as the binding energy
is small. These unique features suggest that the value of the triple
integral mainly depends on the value of $\sum_{|{\bf l}|<k_{F}}(\epsilon_{\bf
k}^{F}+\epsilon_{\bf p}^{F}-\epsilon_{\bf l}^{F}-\epsilon_{F}^{F}+\epsilon_{\bf Q+l-k-p}^{B}-
E-i0^{+})^{-1}$
when $\bf k$ and $\bf p$ are near the two-body continuum threshold; i.e.,
$\left|{\bf k}\right| = \left|{\bf p}\right| = k_{F}$ (so-called
{}``back-to-back scattering'') for $Q=0$, and $\mathbf{k}\simeq\mathbf{p}\simeq
k_{F}\mathbf{Q}/Q$
(so-called ``forward scattering'') for finite $\bf Q$. Based
on these considerations, one can introduce a step function approximation
to simplify the dependence over $\bf k$ and $\bf p$,
\begin{equation}
\frac{1}{\Omega}\sum_{\bf l}\frac{n_{F}({\bf l})}
{\epsilon_{\bf k}^{F}+\epsilon_{\bf p}^{F}-\epsilon_{\bf l}^{F}-\epsilon_{F}^{F}
+\epsilon_{\bf Q+l-k-p}^{B}-E-i0^{+}}\simeq
\begin{cases}
\bar{\Gamma}_{C}, & k_{F}\le\left|{\bf k}\right|,\left|{\bf p}\right|\le\Lambda_{C}\\
0, & \textrm{otherwise}\end{cases},\label{eq:gmb-const}\end{equation}
Here $\bar{\Gamma}_{C}$ is a constant independent of $\bf p$ and $\bf l$.
It is natural to choose $\bar{\Gamma}_{C}$ to be the value for ``back-to-back
scatterings'' or ``forward scatterings'' as
\begin{eqnarray}
\bar{\Gamma}_{C}({\bf Q}=0,E) & = &
\int\frac{d\Omega_{\bf{n_q}}}{4\pi}\frac{d\Omega_{\bf{n_p}}}{4\pi}
\frac{1}{\Omega}\sum_{\bf l} \frac{n_{F}({\bf
l})}{\epsilon^F_F -\epsilon^F_{\bf l} + \epsilon^B_{{\bf l}-k_{F}{\bf n_q}
-k_{F}{\bf n_p}}-E-i0^+}\label{eq:gmb-r-q0}\\
\bar{\Gamma}_{C}({\bf Q}\neq0,E) & = & \frac{1}{\Omega}\sum_{\bf l}
\frac{n_{F}({\bf l})}
{\epsilon^F_F - \epsilon^F_{\bf l} + \epsilon^B_{{\bf Q}+{\bf l}-2k_{F}\frac{\bf Q}{|\bf Q|}} -E-i0^+}\label{eq:gmb-r-qn0}
\end{eqnarray}
where $\bf{n_p}$  and $\bf{n_q}$ are unit vectors.
The effective cutoff $\Lambda_{C}$ specifies the $k$ or $l$ dependence
of the integral in Eq. (\ref{eq:gmb-const}), and is estimated to be
a few $k_{F}$s in our case. In the following we will see that the
specific value of $\Lambda_{C}$ does not enter into the leading order
correction. The triple-momentum integral thus can be simplified as
\begin{eqnarray}
 &  & \left(-\bar{\Gamma}_{C}\right)\left[\frac{2\pi \hbar^2
a}{m_{R}}\frac{1}{\Omega}\sum_{k_{F}<|{\bf k}|<\Lambda_{C}}\frac{1}{\epsilon_{\bf
k}^{F}+\epsilon_{\bf Q-k}^{B}-\epsilon_{F}^{F}-E-i0^{+}}\right]^{2}\\
 & = & \left(-\bar{\Gamma}_{C}\right)\left[\frac{2\pi \hbar^2 a}{m_{R}}
\frac{1}{\Omega}\sum_{\bf k}\left(\frac{\Theta(\left|{\bf k}\right|-k_{F})}{\epsilon_{\bf
k}^{F}+\epsilon_{\bf Q-k}^{B}-\epsilon_{F}^{F}-E-i0^{+}}-\frac{1}{\epsilon_{\bf
k}^{R}}\right)+\left(\frac{1}{\epsilon_{\bf k}^{R}}-\frac{\Theta(\left|{\bf k}\right|-
\Lambda_{C})}{\epsilon_{\bf k}^{F}+\epsilon_{\bf Q-k}^{B}-\epsilon_{F}^{F}-E-i0^{+}}\right)
\right]^{2}\label{eq:gmb-eff1}\\
 & \simeq & \left(-\bar{\Gamma}_{C}\right)
 \left[\frac{2\pi \hbar^2 a}{m_{R}}\left(-\frac{m_{R}}{2\pi \hbar^2
a}+\frac{m_{R}k_{F}}{\hbar^2} O\left(1\right)\right)\right]^{2}\label{eq:gmb-eff2}\\
 & = & \left(-\bar{\Gamma}_{C}\right)\left(1+O(k_{F}a)\right)\label{eq:gmb-
eff}\end{eqnarray}
Here $\Theta(x)$ is the unit step function. In Eq. (\ref{eq:gmb-eff1})
we approximate $E$ to be the mean field solution so that the first
sum is $-\frac{m_{R}}{2\pi \hbar^2 a}$ according to Eq. (\ref{eq:invT-re}).
The second term in Eq. (\ref{eq:gmb-eff1}) is estimated of $\frac{m_{R}k_{F}}{\hbar^2}$,
if $\left(\Lambda_{C}^{2}-k_{F}^{2}\right)m_{R}^{-1}$ is large compared
to the binding energy.

With Eq. (\ref{eq:gmb-eff1}), Eq. (\ref{eq:gmb-eff2}) and Eq. (\ref{eq:bse-gmb}), one arrives at
the main result of this section,
\begin{eqnarray}
-\frac{m_{R}}{2\pi \hbar^2 a} & = & \frac{1}{\Omega}\sum_{k}\left(\frac{1-
n_{F}(k)}{\epsilon_{k}^{F}+\epsilon_{Q-k}^{B}-\epsilon_{F}^{F}-E-i0^{+}}-
\frac{1}{\epsilon_{k}^{R}}\right) - \bar{\Gamma}_{C}({\bf Q}, E)\label{eq:gap-equation-gmb}
\end{eqnarray}
where the effective scattering length $\tilde{a}$ and the function $R$ are
\begin{eqnarray}
\frac{1}{k_{F}\tilde{a}}&=&\frac{1}{k_{F}a}-
\frac{2\pi\hbar^2}{m_{R}k_{F}}\textrm{Re}\bar{\Gamma}_{C}({\bf Q}, E)
\label{es} \\
R&=& \frac{2\pi\hbar^2}{m_{R}k_{F}}\textrm{Re}\bar{\Gamma}_{C}({\bf Q}, E)
\end{eqnarray}
Several remarks about the effective lengths. First $\bar{\Gamma}_{C}$ is usually a complex
number, with its real value leading to dimers' energy shift, and imaginary part relating to dimers'
decay or finite lifetime. Here we only include the real part to obtain the energetics.
Second, effective scattering lengths usually depend on energy and momentum of the pair,
and we have plotted the $R$ function for different $\bf Q$ and $E$ in Fig.\ref{fig7}. In the
$k_Fa\rightarrow0^-$ limit when dimers are shallow, one can neglect the energy
dependence and simply set $R=R({\bf Q},E=E_{th}({\bf Q}))$.
To extrapolate our result to near resonances, we obtain $W_B(Q)$ by solving Eq. (\ref{eq:gap-equation-gmb})
self-consistently, and the results are shown in Fig.\ref{fig7}(b) and Fig.\ref{fig8}.

In the limit of large $k_Fa$, the above ansatz in Eq. (\ref{es}) remains correct,
although the estimate of $\bar{\Gamma}$ should include many other higher order vertex
corrections which are not included in the irreducible diagrams shown in Fig.10 and is therefore more involved.
Nevertheless, we expect that the diagrams in Fig.10 capture the most relevant qualitative aspect
of particle-hole fluctuations.
We therefore extrapolate this result to the unitary limit and
apply it to atoms near interspecies resonance in Sec. III.
Note also that for the dimer energetics in zero- or small-$Q$ channels, the vertex correction appears in the lowest order and is
the most dominating effect of particle-hole fluctuations.
In this limit, the self-energy effect (i.e., the effective mass, the residue of the Green's function)
of the minority particle appears in a higher order in terms of $k_Fa$.
The renormalized chemical potential does not play an important part in the discussion of the dimer; it
gives an overall shift of the two-body continuum as well as the dimer's energy.
We plan to study high-order effects in the future.

\end{widetext}


\begin{thebibliography}{}
\bibitem{TBP}The lifetime of these anomalous molecules will be discussed in a separate paper.
\bibitem{PT}
Following the discussions about the dimer dispersion,
it appears that the weakly interacting state becomes metastable
near the critical line. For Fermi-Fermi mixtures,
this implies a first order phase transition although a proof of
it should involve a more elaborated analysis of fluctuations.
Assuming the transition is of the first order,
beyond the critical line toward the molecular side of the resonance,
the weakly interacting Fermi mixture becomes a supercooled phase in the vicinity of the
critical line or instability point but is globally unstable.
This is also consistent with the first order phase transition between normal polarized electron
liquids or LOFF states and BCS states in superconductors.
For Fermi-Bose mixtures, the issue of the order of phase transition is more subtle
depending on whether dimers form a Fermi surface and whether the atomic
Fermi surface still persists up to this point thus is beyond the scope of
this paper (see also the discussion at the end of the section).
But if the transition is dictated by the emergence of a dimer Fermi surface, then the transition
is likely of the free Fermi gas universality\cite{Powell05}.
\end{thebibliography}
\end{document}